\documentclass[12pt]{article}
\usepackage{amsmath}
\usepackage{amscd}
\usepackage{amssymb}
\usepackage{array}

\begin{document}
\rightline{CERN-TH/2002-373}
\rightline{UCLA/02/TEP/42}
\rightline{hep-th/0212236}

\newcommand{\R}{\mathbb{R}}
\newcommand{\C}{\mathbb{C}}
\newcommand{\Z}{\mathbb{Z}}
\newcommand{\Hb}{\mathbb{H}}

\newcommand{\rE}{\mathrm{E}}
\newcommand{\ii}{\mathrm{i}}
\newcommand{\rSp}{\mathrm{Sp}}
\newcommand{\rSO}{\mathrm{SO}}
\newcommand{\rSL}{\mathrm{SL}}
\newcommand{\rSU}{\mathrm{SU}}
\newcommand{\rUSp}{\mathrm{USp}}
\newcommand{\rU}{\mathrm{U}}
\newcommand{\rF}{\mathrm{F}}
\newcommand{\rGL}{\mathrm{GL}}
\newcommand{\rG}{\mathrm{G}}
\newcommand{\rK}{\mathrm{K}}

\newcommand{\fgl}{\mathfrak{gl}}
\newcommand{\fu}{\mathfrak{u}}
\newcommand{\fsl}{\mathfrak{sl}}
\newcommand{\fsp}{\mathfrak{sp}}
\newcommand{\fusp}{\mathfrak{usp}}
\newcommand{\fsu}{\mathfrak{su}}
\newcommand{\fp}{\mathfrak{p}}
\newcommand{\fso}{\mathfrak{so}}
\newcommand{\fl}{\mathfrak{l}}
\newcommand{\fg}{\mathfrak{g}}
\newcommand{\fr}{\mathfrak{r}}
\newcommand{\fe}{\mathfrak{e}}
\newcommand{\ft}{\mathfrak{t}}
\newcommand{\id}{\relax{\rm 1\kern-.35em 1}}

\vskip 1cm

  \centerline{\LARGE \bf $N=2$ Super-Higgs, $N=1$ Poincar\'e  Vacua}

  \bigskip

   \centerline{\LARGE \bf and Quaternionic Geometry}

 \vskip 1.5cm
\centerline{L. Andrianopoli$^\flat$, R. D'Auria$^\sharp$
,  S.
Ferrara$^{\flat \star}$ and
 M. A. Lled\'o$^\sharp$ }
%\vskip 5mm
%\centerline{\small e-mail: Laura.Andrianopoli@cern.ch,  dauria@polito.it, }
%\centerline{\hskip 1.2cm \small Sergio.Ferrara@cern.ch, lledo@athena.polito.it}
\vskip 1cm

\centerline{\it $^\flat$ CERN, Theory Division, CH 1211 Geneva 23,
Switzerland.}

\bigskip

\centerline{\it $^\sharp$ Dipartimento di Fisica, Politecnico di
Torino,} \centerline{\it Corso Duca degli Abruzzi 24, I-10129
Turin , Italy  and  } \centerline{\it   INFN, Sezione di Torino,
Italy. }

\bigskip

\centerline{$^\star$ \it INFN, Laboratori
Nazionali di
Frascati, Italy and}
\centerline{\it Department of Physics and Astronomy, University of
California,} \centerline{\it  Los Angeles, USA}

\vskip 3cm

\begin{abstract}
In the context of $N=2$ supergravity we explain the occurrence of partial super-Higgs with vanishing vacuum energy and
moduli stabilization  in a model suggested by superstring compactifications on type IIB orientifolds with 3-form fluxes.

The gauging of axion symmetries of the quaternionic manifold, together with the use of degenerate symplectic sections
for special geometry, are the essential ingredients of the construction.

\end{abstract}

 \vfill\eject

\section{Introduction}

A general and challenging problem in effective superstring theories described by supergravity lagrangians is to 
understand the nature of spontaneous supersymmetry breaking in a model independent fashion \cite{ps} - \cite{adfl1}.

Superstring vacua obtained by IIB orientifolds with three-brane fluxes turned on \cite{drs} - \cite{fm}
 offer suitable examples where the spontaneous
breaking of $N=4 \to N=0$ supersymmetry can occur by stepwise breaking of supersymmetry $N \to N-1$ 
with vanishing vacuum energy.

Although quantum corrections may spoil this mechanism \cite{bbhl} (especially if the true vacuum has no supersymmetry), 
the hope remains that this mechanism may create a hierarchy of scales, as was suggested in the old no-scale supergravity models
 \cite{cfkn,elnt}.
We will indeed show that such vacua, giving partial breaking of supersymmetry, are generalized no-scale models.

In extended supergravity the no-scale structure, which results in a positive potential with vacua exhibiting flat directions,
crucially depends on the gauge group at work.

The model we will consider here has the  property that the scalar manifold of hypermultiplets has some translational isometries 
corresponding to shifts in the ``axion scalars'' directions.
This in turn implies that a ``triangular parametrization'' can be given of the manifold, where the axions are 
contained in the off-diagonal block of the vielbein one form \cite{fgp,tz,zino}.

We consider the $N=2$ theory based on the following non linear $\sigma$-model
$${\mathcal M} = {\mathcal M}_V \times {\mathcal M}_Q = \frac{\rSU(1,1)}{\rU(1)} \times \frac{\rSU(2,2)}{\rSU(2)\times \rSU(2)\times
 \rU(1)}$$
where $ \frac{\rSU(1,1)}{\rU(1)}$ is the special K\"ahler manifold which corresponds, in a certain (non-degenerate) choice
of special coordinates, to a linear prepotential 
\begin{equation}
F(X) = \ii X^0X^1; \, f(z) = \ii z \mbox{ with } z = X^1/X^0 .
\label{standard}
\end{equation}

The two dimensional quaternionic manifold ${\mathcal M}_Q$ 
\footnote{We call $n_H$ the (quaternionic) dimension of a quaternionic manifold, where $4n_H$ is its real dimension.
$n_H$ is the number of hypermultiplets of the given $N=2$ theory.}
 is also K\"ahler and it has four translational isometries, corresponding 
to the decomposition
\begin{equation}
 \rSU(2,2) \to \rSL(2,\C) \times\rSO(1,1) \label{decomp}
\end{equation}
under which 
$$ \fsu(2,2) \to \fsl(2,\C) +\fso(1,1) + {\bf 4}^++ {\bf 4}^-.$$ 

In the gauged version of the model we use the two vectors (the graviphoton plus one matter vector) to  gauge two of
the four  translational isometries ${\bf 4}^+$ of $\rSU(2,2)$. More precisely, since ${\bf 4}^+$ is a Lorentzian four-vector, one
has to gauge two out of the three spatial components of ${\bf 4}^+$.

This model is an $N=2$ truncation \cite{adfl2,adfl3} of a type IIB 
$T^6/\Z_2$ orientifold with fluxes. 
Here the two remainig fluxes are described by two gauge coupling constants.

A crucial point of this gauging is that it must be formulated in a duality basis \cite{gz}
 for special geometry which has a degenerate holomorphic
section $X^\Lambda(z)$ (and where no function $F(X)$ exists) \cite{cdfv,fgp}, 
that is in a basis non locally related to the ``standard'' one (\ref{standard}).
This is because otherwise it would be impossible to break $N=2 \to N=1$ since $N=2$
would be either unbroken or completely broken \cite{cgp1,cgp2}.
However, the no-go theorem \cite{cgp2} is based on the applicability of the $N=2$ tensor calculus \cite{dlv}, 
which precisely fails for those symplectic
bases for which no prepotential function $F(X)$ exists.

The role of different choices of symplectic embeddings of duality symmetries has been investigated in recent time 
\cite{adfl2,adfl3,hull}

The symplectic basis to be chosen, to realize the partial breaking of supersymmetry, is the one considered in reference \cite{fgp}
and further extended to the case of $\frac{\rSU(1,1+n)}{\rSU(1+n) \times \rU(1)}$ in \cite{dflv}.
In this basis we have
$X^\Lambda = (X^0 , -\ii X^0 )$ and $F_\Lambda = (\ii X^1 , X^1)$, so that in the $z$ coordinate $z=X^1/X^0$ the symplectic section is:
\begin{equation}
X^\Lambda =(1,-\ii) , \, F_\Lambda =(\ii z , z).
\label{basis}
\end{equation}
Correspondingly, the K\"ahler potential of the $\frac{\rSU(1,1)}{\rU(1)}$ special manifold is:
\begin{equation}
{\mathcal K} = - {\rm log}[\ii(\bar X^\Lambda F_\Lambda - X^\Lambda \bar F_\Lambda )]= - {\rm log}[-2(z+\bar z)] ; \quad \, \Re z < 0
\end{equation}   

The reason for this choice of duality basis comes from the embedding of the type IIB vectors in the duality group \cite{fp1,kst,dfv}.
This requires the choice of a basis in which the $\theta$ term (given by $\Re {\mathcal N}_{\Lambda\Sigma}$ in special geometry
\cite{abcdffm}) is proportional to the axion contained in the vector multiplet. Indeed, in our basis the kinetic vector matrix is
${\mathcal N} = \ii z \id$, which has the required property.

 \section{Quaternionic manifolds and axion symmetries}

The manifold $G/H =\frac{\rSU(2,2)}{\rSU(2)\times \rSU(2)\times \rU(1)}$ can be parametrized, according to the decomposition
(\ref{decomp}), through a coset element which is
the following $\rSU(2,2)$ $4\times 4$ matrix
\begin{equation}
L = \begin{pmatrix}E & -BE^{-1} \cr 0 & E^{-1}\cr\end{pmatrix}
\end{equation}
where $E$ is an element of $\frac{\rSL(2,\C)}{\rSU(2)}\times \rSO(1,1)$ 
\begin{equation}
E = e^0 \id + e^i \sigma^i \quad  (e^0, e^i \in \R;\; i =1,2,3;\; \sigma^i\, \mbox{are Pauli matrices})
\end{equation} 
with $e^0 >0$ and ${\rm det}(E) \equiv e^2 = (e^0)^2-e^ie^i >0$,
and $B$ is an antihermitian matrix
\begin{equation}
B = \ii b^0 \id + \ii b^i \sigma^i \quad (b^0, b^i \in \R).
\end{equation}

Its left invariant one-form $\Gamma = L^{-1} dL $, satisfying the Maurer--Cartan equation
$d\Gamma + \Gamma \wedge \Gamma =0$  is an $\fsu(2,2)$ matrix
\begin{equation}
L^{-1} dL = \begin{pmatrix}E^{-1} dE & -E^{-1}dBE^{-1} \cr 0 &EdE^{-1}\cr\end{pmatrix}. \label{maurer}
\end{equation}

In order to extract from it  the expressions for the $H$-connection and vielbein of $G/H$, let us compare 
 (\ref{maurer}) 
with the general form of an arbitrary element of the $\fsu(2,2)$ Lie algebra, satisfying ${\rm Tr}A =0$; $\eta A^\dagger \eta = -A $
 with 
 $\eta = \begin{pmatrix}0 & \id \cr \id & 0\cr\end{pmatrix}$,
\begin{equation}
A=\begin{pmatrix} p \id + (q_1 + \ii q_2 )^i  \sigma^i & \ii (r \id + s^i \sigma^i) + \ii (t\id +u^i\sigma^i) \cr
 \ii (r\id + s^i \sigma^i) - \ii (t\id +u^i\sigma^i) & -p \id - (q_1 - \ii q_2 )^i\sigma^i \cr
\end{pmatrix}
\end{equation}
where the 15 elements $p,q^i_1, q^i_2, r,s^i, t, u^i$ are all real.
The maximal compact subalgebra $\fsu(2)\times \fsu(2)\times \fu(1)$ is given by the antihermitean part of $A$:
\begin{equation}
h=\frac 12 (A-A^\dagger) =\begin{pmatrix}  \ii q_2^i  \sigma^i & \ii (r \id + s^i \sigma^i)  \cr
 \ii (r\id + s^i \sigma^i) &  \ii q_2^i\sigma^i \cr
\end{pmatrix}
\end{equation}
while the generators of the coset $\rSU(2,2)/[\rSU(2)\times \rSU(2) \times \rU(1)]$ are in the hermitean part of $A$
\begin{equation}
k=\frac 12 (A+A^\dagger) =\begin{pmatrix}
p \id + q_1^i  \sigma^i &  \ii (t\id +u^i\sigma^i) \cr
 - \ii (t\id +u^i\sigma^i) & -p \id - q_1^i\sigma^i \cr
\end{pmatrix}
\end{equation}

By comparison, noting that for the left-invariant form $\Gamma$ of (\ref{maurer}) we have:
\begin{eqnarray}
&&r=t \,; \quad s^i = u^i  \nonumber\\
&& p \id + q_1^i  \sigma^i = \frac 12 (E^{-1} dE + \mbox{ h. c.}) \nonumber\\
&& \ii q_2^i \, \sigma^i = \frac 12 (E^{-1} dE - \mbox{ h. c.}) \nonumber\\
&&2 \ii (r \id + s^i \sigma^i) = -E^{-1} dB E^{-1}
\end{eqnarray}
we finally obtain the decomposition of (\ref{maurer}) in vertical (G/H) and horizontal (H) components:
\begin{equation}
(L^{-1} dL)_H = \begin{pmatrix} \omega_1 & \omega_0 + \omega_2 \cr \omega_0 + \omega_2 & \omega_1\cr
\end{pmatrix} = \Omega_{\fsu(2)\times \fsu(2)\times \fu(1)}
\end{equation}
with
\begin{eqnarray}
\omega_0 = \omega_0 \id  & =&-\frac \ii {2(e^2)^2}[((e^0)^2 +e^ie^i)db^0 -2e^0e^i db^i]\id \nonumber\\
\omega_1 = \omega_1^i \sigma^i& =& -\frac \ii {e^2} \epsilon^{ijk} e^jde^k  \sigma^i \nonumber\\
\omega_2 = \omega_2^i \sigma^i& =& -\frac \ii {2(e^2)^2}[ -2e^0e^i db^0+ (e^2\delta^{ij} +2e^ie^j)db^j] \sigma^i
\end{eqnarray}
 and
\begin{equation}
(L^{-1} dL)_{G/H} = \begin{pmatrix} V_0+V_1
 & \omega_0 + \omega_2 \cr -\omega_0 - \omega_2 & -V_0-V_1\cr
\end{pmatrix}= {\mathcal V}
\end{equation}
with
\begin{eqnarray}
V_0   & =&\frac 1 {e^2}[e^0 de^0 -e^ide^i]\id \nonumber\\
V_1  & =&\frac 1 {e^2}[e^0 de^i -e^ide^0]\sigma^i .
\end{eqnarray}
The vielbein one-form has the off-diagonal components which are given in terms of the off-diagonal components of the H-connection.
As we will see in the next section, this peculiar fact is at the origin of the no-scale structure of the theory.

The kinetic energy term is given by 
 \begin{equation}
h_{uv} d q^u d q^v = \frac 12 {\rm Tr} \left[\left(L^{-1} dL\right)_{G/H} \cdot
\left(L^{-1} dL\right)_{G/H}\right].\label{kinscal}
\end{equation}
In the gauged theory the differential $dq$ is replaced by $Dq$, which implies that $dL \to DL$ in (\ref{kinscal}). 
In the following we will denote by ${\hat \omega}$, ${\hat{\mathcal V}}$  the gauged connection and vielbein in $L^{-1}DL$
related to the corresponding ungauged objects of (\ref{maurer}).
\bigskip

Let us conclude this paragraph by noticing that another (one dimensional) quaternionic manifold, 
$\frac{\rUSp(2,2)}{\rUSp(2) \times \rUSp(2)}$, can be obtained from the former if we further impose on $L$ to be symplectic, {\em i.e.}
$$L^T \Omega L =\Omega \; ,\quad \mbox{with } \Omega = \begin{pmatrix}0&\epsilon \cr \epsilon &0 \cr \end{pmatrix}.$$ 
It is straightforward to show that this sets $e^i = b^0=0$ so that
 \begin{eqnarray}
L& =& \begin{pmatrix}a\id & -\ii a^{-1}b^i\sigma^i \cr 0 & a^{-1}\id 
\cr\end{pmatrix}
\nonumber\\
L^{-1} dL &=& \begin{pmatrix}a^{-1} da\id & -\ii a^{-2}db^i \sigma^i \cr 0 &-a^{-1} da\id \cr \end{pmatrix} \nonumber\\
(L^{-1} dL)_H &=& \begin{pmatrix}0 & -\frac\ii {2a^{2}}db^i \sigma^i \cr -\frac\ii {2a^{2}}db^i \sigma^i  &0\cr\end{pmatrix} =
\Omega_{\fsu(2)_R}\nonumber\\
(L^{-1} dL)_{G/H} &=& \begin{pmatrix}a^{-1} da \id  & -\frac\ii {2a^{2}}db^i \sigma^i \cr \frac\ii {2a^{2}}db^i \sigma^i  & -a^{-1} da \id
\cr\end{pmatrix} =
{\mathcal V} \label{parametr}
\end{eqnarray}
We note that also in this case the off-diagonal component of ${\mathcal V}$ 
are related to the $\fsu(2)$ connection that here coincides with the R-symmetry connection $\omega^i_{R}= \frac{db^i}{a^2}$.
This is the quaternionic manifold underlying the simplest example of $N=2$ gauged supergravity with partial breaking of supersymmetry
considered in the literature \cite{cgp3,fgp}
\footnote{
Note that, in order to compare (\ref{parametr}) with reference \cite{fgp},  we have to set, for the coordinate $b_0$ of \cite{fgp}, 
$b_0=a^2$.}.

\section{Gauging quaternionic isometries; scalar potential and masses}

In absence of non abelian gauging, $N=2$ supergravity predicts a scalar potential of the form \cite{abcdffm}
\begin{equation}
V = 4 h_{uv} k^u_\Lambda k^v_\Sigma\bar X^\Lambda X^\Sigma {\rm e}^{\mathcal K} + U^{\Lambda\Sigma}
P^i_\Lambda P^i_\Sigma 
-3 P^i_\Lambda\bar X^\Lambda
 P^i_\Sigma X^\Sigma {\rm e}^{\mathcal K}\,, \quad (i=1,2,3). \label{potential}
\end{equation}

The first term is the contribution of the hyperinos variation, the second term is the contribution of the 
gauginos and the third of the gravitinos.

The matrix $U^{\Lambda\Sigma}$ is given, by special geometry, to be 
\begin{equation}
U^{\Lambda\Sigma} ={\rm e}^{\mathcal K} {\mathcal D}_i X^\Lambda{\mathcal D}_{\bar\jmath}\bar X^\Sigma g^{i\bar\jmath}=
-\frac 12 (\Im {\mathcal N}_{\Lambda\Sigma})^{-1} - {\rm e}^{\mathcal K}\bar X^\Lambda X^\Sigma 
\end{equation}
In the case at hand, the symplectic section (\ref{basis}) gives, for the kinetic matrix,
  ${\mathcal N}_{\Lambda\Sigma}= \ii z \id$, so that
 $-\frac 12 (\Im {\mathcal N}_{\Lambda\Sigma})^{-1}=2{\rm e}^{\mathcal K}\delta_{\Lambda\Sigma}$ and we get
\begin{equation}
U^{\Lambda\Sigma} P^i_\Lambda P^i_\Sigma = -\frac 1 {2 (z+\bar z)}\left(
P^i_1 P^i_1 + P^i_2 P^i_2\right)=
P^i_\Lambda P^i_\Sigma \bar X^\Lambda X^\Sigma {\rm e}^{\mathcal K}
\label{equiv}
\end{equation}
From (\ref{equiv}), the second and third terms in (\ref{potential}) 
together give a negative contribution to the potential:
\begin{equation}
-2 P^i_\Lambda \bar X^\Lambda P^i_\Sigma X^\Sigma {\rm e}^{\mathcal K} = \frac 1 {(z+\bar z)}
\left(P^i_1 P^i_1 + P^i_2 P^i_2\right).
\end{equation}
The above term can be further simplified by noting that, when gauging axion symmetries (this actually implies, 
in the parametrization chosen, that all components of the $\rSU(2)$ connections are $b$-independent), we have \cite{abcdffm,mic,agata}
\begin{equation}
P^i_\Lambda = (\omega^i_R)_u k^u_\Lambda
\label{simplep}
\end{equation}
where $ (\omega^i_R)_u$ is the component of the R-symmetry $SU(2)$ connection $\omega_R^i\equiv 2\ii(\omega_1^i+\omega_2^i)$.

The scalar potential then becomes
\begin{equation}
V =\frac 14 {\rm e}^{\mathcal K}\left[ \left(4 h_{b_1 b_1} - 2 \omega^i_{b_1} \omega^i_{b_1}\right)g_1^2 +
 \left(4 h_{b_2 b_2} - 2 \omega^i_{b_2} \omega^i_{b_2}\right)g_2^2 \right]
, \label{potentialsimpl}
\end{equation}
where we have taken $k^{b_i}_0 =\frac{g_1}2\, \delta ^i_1  $, $k^{b_i}_1 =\frac{g_2}2\, \delta ^i_2  $.

By virtue of the actual form of $(L^{-1}dL)_{H}$ and $(L^{-1}dL)_{G/H}$, we notice that the $\omega_2$ contributions
exactly cancel in $V$, with the $\omega_0$ terms left, so that
\begin{equation}
V=\frac 2{(e^2)^4} {\rm e}^{\mathcal K}  e_0^2 (e_1^2 g_1^2 + e_2^2 g_2^2)
\end{equation}
which gives
\begin{equation}
\frac{\partial V}{\partial e_1} = \frac{\partial V}{\partial e_2}=0 \quad \Rightarrow \quad e_1 =e_2 =0
\end{equation}
while $z$, $e_0$ and $e_3$ are flat directions.

From the normalization of the kinetic term of the hypermultiplets, setting $e_1 =e_2 =0$, so that $e^2 = e_0^2 - e_3^2$, we have, for the two scalars $e_1, e_2$ 
  \begin{equation}
h_{e_i e_j }de^i de^j = {\rm Tr}(V_1\cdot V_1) = \frac 2{(e^2)^2}
e_0^2 \left((de_1)^2 + (de_2)^2\right) \, \quad (i,j=1,2)
\end{equation}
so that the scalar masses are $ m_i^2 = {\rm e}^{\mathcal K} \frac{g_i^2}{(e^2)^2}$.

The gravitino mass matrix is given by \cite{abcdffm}
\begin{equation}
S_{AB} =\frac \ii 2 {\rm e}^{\frac{\mathcal K}2}P^i_\Lambda X^\Lambda (\sigma^i\epsilon )_{AB}=\frac \ii 4 {\rm e}^{\frac{\mathcal K}2}
\left[ g_1 (\omega^i_R)_{b_1} -\ii g_2 (\omega^i_R)_{b_2}\right](\sigma^i\epsilon )_{AB}.
\end{equation}
From the explicit form of $(\omega^i_R)_b = 2\ii(\omega^i_2)_b$ 
  we read (at $e_1=e_2=0$) 
\begin{equation}
(\omega_R)_{b_i} =- \frac 1{(e^2)^2} (e_0^2 - e_3^2) (\sigma_i \epsilon)  \, , \quad (i =1,2)
\end{equation}
so that,  recalling that
$\sigma_1 \epsilon = -\sigma_3$,  $\sigma_2 \epsilon = \ii \id$
($\epsilon = \ii \sigma_2$) and that the physical gravitino masses are given by the eigenvalues of $2S_{AB}$
\begin{equation}
S_{AB} = -\frac \ii 2\frac{ {\rm e}^{\frac{\mathcal K}2}}{e^2}
\begin{pmatrix} \frac{g_2 -g_1}2 & 0 \cr 0 & \frac{g_1 + g_2}2 \cr \end{pmatrix}
\end{equation}
Unbroken $N=1$ supersymmetry requires $|g_1|=|g_2|= g$, in which case 
the hyperscalars and the gravitino masses are just equal to each other
\begin{equation}
m^2 = \frac  {{\rm e}^{\mathcal K}}  {(e^2)^2}g^2
\end{equation}

The vector bosons masses come from the gauge covariant derivative $Db^i$ ($i=1,2$) in the hypermultiplets (gauged) kinetic term
\begin{eqnarray}
h_{b_ib_j} Db^i Db^j&=&  -{\rm Tr}(\hat \omega_2 \cdot \hat \omega_2)
= \frac 1 {2(e^2)^2}  \left[(db^1 + \frac 12 g_1 A^0)^2+ (db^2 + \frac 12 g_2 A^1)\right]^2 \nonumber\\
&=& \cdots + \frac 1{8(e^2)^2}[g_1^2 (A^0)^2 + g_2^2 (A^1)^2].
\end{eqnarray}
This term has to be confronted with the vectors kinetic term in the lagrangian:
\begin{equation}
\Im {\mathcal N}_{\Lambda\Sigma}F^\Lambda_{\mu\nu}F^{\Sigma \mu\nu}= \Re z [(F^0_{\mu\nu})^2 +(F^1_{\mu\nu})^2]
\end{equation}
where we have chosen the field strenghts normalization $F_{\mu\nu}^\Lambda =\frac 12 (\partial _\mu A_\nu^\Lambda - \partial _\nu A_\mu^\Lambda)$.
Using the fact that $\Re z=- \frac 14 {\rm e}^{-{\mathcal K}}$,  the standard contribution $F^2 + \frac 12 A^2$
is finally got with the vector (squared) masses
\begin{equation}
m_i^2 = \frac{{\rm e}^{\mathcal K}}{(e^2)^2} g_i^2.
\end{equation}

Note that this result is also true in the one dimensional case of \cite{fp2}, since the number of vector bosons and charged axions
is the same in the two theories.

\bigskip

We just note that in the case of the one dimensional quaternionic manifold
$\frac{\rUSp(2,2)}{\rUSp(2)\times \rUSp(2)}$ \cite{cgp2,fgp} the $\omega_0$ term is absent so that
\begin{equation}
4h_{b_ib_i} -2 \omega^x_{b_i} \omega^x_{b_i} =0
\end{equation}
giving $V\equiv 0$, instead of being $\frac 8{(e^2)^4} e_0^2 e_e^2$ as in the actual case, which leads to $V>0$.

\section{Relation to $N=2$ warped compactifications}

The model considered here is the bulk massless sector of the effective theory of the type IIB orientifold \cite{fp1,kst}
with $N=4$ supersymmetry partially broken to $N=2$ and then to $N=0$.
Technically this result is obtained by integrating out two of the four gravitino multiplets, 
by assuming that $m_3, m_4 >>m_1, m_2$.
In particular, let us start with the $\rSO(6,6) / \rSO(6)\times \rSO(6)$ manifold with coordinates $g_{IJ}=g_{JI}$, 
$b_{IJ} = -b_{JI}$ ($I,J=1,\dots , 6$). If we use complex coordinates $I=(i,\bar\jmath )$ ($i,\bar\jmath=1,2,3$), the $N=4 \to N=3$
truncation corresponds to keep all $(g_{i\bar\jmath}, b_{i\bar\jmath})$ components (and not $g_{ij}$, $b_{ij}$).
A further reduction to $N=2$ corresponds to retain only {\em e.g.}
($g_{1,\bar 1}, g_{i\bar\jmath}$), ($b_{1,\bar 1}, b_{i\bar\jmath}$) with $i,\bar\jmath =2,3$.
Following the notations of \cite{dfv}, the two massive gravitino multiplets which break $N=4 \to N=3 \to N=2$
correspond to fluxes $f_{123}, f_{\bar 1 23}$. The residual supersymmetry is then broken $N=2 \to N=1 \to N=0$ by the 
fluxes $f_{1\bar 2 3}, f_{\bar 1 \bar 2 3}$.

It is interesting that, at each stage of partial breaking, we get an effective no-scale supergravity model,
with a rather simple geometric structure.
The coset described by the $E$ coordinates corresponds to the metric moduli of the torus, together with $\Re z$.
The five fields $b_0,b_i, \Im z$ correspond to the R-R axions which are retained in the $N=2$ truncation.

This model can also be obtained by starting with a $N=3$ effective theory with 3 vector multiplets \cite{fp1}
and integrating out a long $N=2$ spin $3/2$ multiplet $\left((\frac 32), 4(1), (5+1) (\frac 12), 4(0)\right)$.
One sees that the remaining massless degrees of freedom correspond exactly to one vector multiplet and two hypermultiplets 
as well as the $N=2$ graviton multiplet.

The $N=3$  manifold $\frac {\rSU(3,3)}{\rSU(3)\times \rSU(3) \times \rU(1)}$ gets reduced to 
$\frac {\rSU(1,1)}{ \rU(1)}\times \frac {\rSU(2,2)}{\rSU(2)\times \rSU(2) \times \rU(1)}$
where $4$ of the $9$ axions of the $N=3$ manifold have been eaten by the 4 vectors of the massive spin-$3/2$ multiplet
and other 4 metric moduli are the scalar superpartners of the massive gravitino.

It is obvious that the present example can be generalized to include, besides the abelian interactions 
gauging the axion isometries, an arbitrary interaction with $n$ Yang--Mills supermultiplets \cite{grana,fm}, 
with a coset structure \cite{fp2,dflv}
$$\frac {\rSU(1,1+n)}{\rU(1)\times \rSU(1+n)}\times \frac {\rSU(2,2+n)}{\rSU(2)\times \rSU(2+n) \times \rU(1)}$$
where $n={\rm dim}\,G_{YM}$.

Note that this structure differs from the one considered in \cite{fgpt}. The vector multiplets part is the same as the one 
described in \cite{zino}.

The vacuum state can at most break $G_{YM} \to CSA$ so that $n$ is reduced to the rank of $G_{YM}$.

In the supergravity framework the Yang--Mills part corresponds to the D3-brane contribution
to the four-dimensional effective theory. \cite{fp1,kst,grana,fm}.

\section{Comparison with Calabi--Yau compactifications and other models}

The present theory can be compared to other effective 
$N=2$ supergravity theories  considered in the literature \cite{tv} - \cite{louis}, \cite{lm} - \cite{hl}, \cite{bbhl,agata,bd}.

For example, in the context of Calabi--Yau compactifications in type IIB superstring, turning on fluxes corresponds 
to gauge axion symmetries of the quaternionic manifold, which is obtained by c-map \cite{cfg} of  
some special K\"ahler manifold \cite{mic,agata}.

In the case of a quaternionic geometry obtained by c-map of a special geometry with cubic prepotential,
 it was shown by Taylor and Vafa \cite{tv} that $V>0$.
This result is true for an arbitrary special K\"ahler geometry
as appropriate in type IIB Calabi--Yau compactifications.
A positive potential was obtained at the $N=2$ level in \cite{agata}
in the particular simple case of the two dimensional quaternionic manifold $G_2 /\rSO(4)$.

However the basic relation which exists in this case is \cite{agata}
\begin{equation}
P^x_\Lambda P^x_\Sigma = h_{uv} k^u_\Lambda k^v_\Sigma
\end{equation}
which then implies
\begin{equation}
V = \left(U^{\Lambda \Sigma} + {\rm e}^{\mathcal K}\bar X^\Lambda  X^\Sigma\right)P^x_\Lambda P^x_\Sigma = -\frac 12 \left(
\Im {\mathcal N}_{\Lambda\Sigma}\right)^{-1} P^x_\Lambda P^x_\Sigma
\end{equation}

The problem with this expression is that it never vanishes unless $P^x_\Lambda =0$ since $-\Im {\mathcal N}_{\Lambda\Sigma}>0$.

This is partly a consequence of the no-go theorem of \cite{cgp2} which implies that, in presence of non-degenerate sections $X^\Lambda$ of
special geometry, either $N=2$ is unbroken or it is broken to $N=0$. This point was stressed in reference \cite{mayr}.

Our examples evade this no-go theorem because we use degenerate sections for the vector multiplets geometry.
In our case the $\rSU(1,1)$ acting on the two vectors mixes electric with magnetic field strenghts.
This is related to the fact that in the embedding $\rSU(1,1) \times \rSU(2,2) \subset \rSO(6,6)$ but the $\rSU(1,1)$ factor 
is not a subset of the electric
$\rGL(6) \subset \rSO(6,6)$, which is the maximal subgroup  acting linearly on the vector potentials
 of the parent $N=4$ theory \cite{adfl3, dfv}.

\section{Quantum corrections to the cosmological constant}

In the present models, giving partial super-Higgs around Minkowski vacuum, we can make some discussion on the one-loop 
corrections to the cosmological constant.

Let us remind that the quartic, quadratic and logarithmic divergent parts, in any field theory, are respectively controlled
by the following coefficients \cite{zum,dewitt}
\begin{equation}
a_k = \sum_J (-1)^{2J}(2J+1)m_J^{2k} \; ,\quad (k=0,1,2).
\end{equation}
While $a_0=0$ in any spontaneously broken supersymmetric theory (quartic divergence), the vanishing of $a_1$, $a_2$ is 
model-dependent \cite{zum}

However, under some mild assumption, in any spontaneously broken $N$-extended theory where the partial super-Higgs $N \to N-1$
is permitted, the following main formulae are true
\begin{equation}
a_k = \sum_J (-1)^{2J}(2J+1)m_J^{2k} =0 \; ,\quad 0\leq k<N.
\end{equation}
This relation follows from the fact that a non vanishing contribution to the vacuum energy
 must be proportional to $\Pi_{i=1}^{N} \frac{m_i^2}{M_P^2} M_P^{4}$.

This is the leading behavior in the variable $X=\Pi_{i=1}^{N} \frac{m_i^2}{M_P^2}$, at least if we assume analyticity in this variable.

From  the above we conclude that there is only a finite correction in $N=3,4$ models, while there is a logarithmic correcton to the 
$N=2$ models and a quadratic correction to the $N=1$ models.

Note however that in the Scherk--Schwarz $N=8$ models \cite{ss} the same formulae were true but for $0\leq k<\frac N2$.
This is because in that case the gravitino masses were pairwise degenerate so that the hypotesis of partial breaking $N \to N-1$ was
invalid.

In our model all fermions helicities have masses $|g_1 \pm g_2|/2$ while the two vectors and the two massive scalars have squared 
masses $g_1^2$, $g_2^2$.\footnote{The masses are given in units of the moduli dependent factors
$\frac
{
{\rm e}^
{\frac{\mathcal K}2}}{e^2}$.}

Since we have 8 helicities with mass $|g_1 + g_2|/2$ and 8 helicities with mass $|g_1 - g_2|/2$, we find
$
2(g_1+g_2)^2 + 2(g_1-g_2)^2 = 4 g_1^2+4g_2^2
$
which is the same as the bosonic contribution $3g_1^2 +3g_2^2 +g_1^2+g_2^2$.
Therefore we have $\sum_J (-1)^{2J} (2J+1)m_J^2 =0$.
(In the model of \cite{cgp2} the bosonic and fermionic contributions were separately equal $3(g_1^2 + g_2^2)$,
 so that ${\rm STr}{\mathcal M}^2=0$ also in that case).
Note that for $|g_1|=|g_2|$ $N=1$ is unbroken, three spin 1/2 fermions  are massless and three have masses $|g|$.
They join three massless chiral multiplets, one massive gravitino multiplet and an extra massive chiral multiplet.

The quartic mass formula
\begin{equation}
 \sum_J (-1)^{2J} (2J+1)m_J^4 = {\rm STr} {\mathcal M}^4
\end{equation}
gives a non vanishing result. It is positive 
$${\rm STr} {\mathcal M}^4= 3 (g_1^2-g_2^2)^2\frac{{\rm e}^{2{\mathcal K}}}{(e^2)^4}= 48m_1^2m_2^2.
$$
In the model of \cite{cgp2} it is instead ${\rm STr} {\mathcal M}^4= 36m_1^2m_2^2$.

\section{Conclusions}
In this paper we have considered a simple $N=2$ lagrangian which correctly reproduces a no-scale extended
supergravity model with vanishing vacuum energy and moduli stabilization.

The crucial ingredients, which appear to be quite general and not only inherent to the case under investigation, are two.
It is necessary to gauge some translational isometries of the quaternionic manifold, in order to be left with a positive semidefinite 
scalar potential giving partial super-Higgs. In this respect,
the use of a degenerate symplectic section for special geometry is further needed, in order to escape the no-go theorem and allow stepwise
supersymmetry breaking $N=2\to N=1$.

A property of such models is that a non trivial moduli space exists in each broken phase.
This model is supposed to describe the bulk sector of a type IIB orientifold in presence of fluxes.
The three-form fluxes are proportional to $g_1 \pm g_2$, where $g_1,g_2$ are the two gauge couplings of the theory.

This model can be generalized to include Yang--Mills interactions by adding $N=2$ vector multiplets and $N=2$ hypermultiplets in the
adjoint representation of some compact Lie group.

The special geometry relevant for this generalized case, with degenerate symplectic sections, has been described in \cite{dflv},
and a natural parametrization of the quaternionic manifold also exists.

It is also natural to extend this analysis to manifolds which are not symmetric spaces as in the present case, but 
still have some abelian isometries to be gauged \cite{louis}.
We expect some of the properties shown here will also apply, under suitable assumptions, to these more general cases.

\section*{Acknowledgements}
We acknowledge interesting discussions with G. Dall'Agata and R. Rattazzi.
 Work
supported in part by the European Community's Human Potential
Program under contract HPRN-CT-2000-00131 Quantum Space-Time, in
which L. A.,  R. D. and M. A. Ll. are associated to Torino
University. This work has also  been supported by the
D.O.E. grant DE-FG03-91ER40662, Task C.

\end{document}